\documentclass[amsmath,amssymb]{revtex4}
\usepackage{graphicx}
\usepackage{amscd,amsmath,amsthm,amsfonts,amssymb}
\def\PT{$\cal{PT}$}
\def\[{\begin{equation}}
\def\]{\end{equation}}

\begin{document}
\title{New families of non-parity-time-symmetric complex potentials with all-real spectra}
\author{Bijan Bagchi$^a$ and Jianke Yang$^b$}
\affiliation{$^a$Department of Physics, Shiv Nadar University, Gautam Buddha Nagar, Uttar
Pradesh 201314, India \\
$^b$Department of Mathematics and Statistics, University of Vermont, Burlington, VT 05405, USA}

\begin{abstract}
New families of non-parity-time-symmetric complex potentials with all-real spectra are derived by the supersymmetry method and the pseudo-Hermiticity method. With the supersymmetry method, we find families of non-parity-time-symmetric complex partner potentials which share the same spectrum as base potentials with known real spectra, such as the (complex) Wadati potentials. Different from previous supersymmetry derivations of potentials with real spectra, our derivation does not utilize discrete eigenmodes of base potentials. As a result, our partner potentials feature explicit analytical expressions which contain free functions. With the pseudo-Hermiticity method, we derive a new class of non-parity-time-symmetric complex potentials with free functions and constants, whose eigenvalues appear as conjugate pairs. This eigenvalue symmetry forces the spectrum to be all-real for a wide range of choices of these functions and constants in the potential. Tuning these free functions and constants, phase transition can also be induced, where conjugate pairs of complex eigenvalues emerge in the spectrum.
\end{abstract}

\maketitle

\section{Introduction}
Construction of potentials with a given spectrum in the linear Schr\"odinger operator has a long history in quantum mechanics \cite{Infeld_1951, Bhatt_1962, Levai_1989, Cooper_1995}. For real potentials, the Schr\"odinger operator is Hermitian, and thus its spectrum is all-real. In 1998, Bender and Boettcher \cite{Bender1998} showed that, if the potential is complex but parity-time (\PT) symmetric, i.e., $V^*(-x)=V(x)$, then the spectrum of the Schr\"odinger operator can still be all-real, unless a phase transition occurs where complex conjugate pairs of eigenvalues appear. This observation opened a new area of inquiry in the realm of non-Hermitian quantum mechanics \cite{Mostafazadeh_quantum, Moise_2011, Brody_2016}. Since then, real spectra in many \PT-symmetric potentials have been analytically demonstrated (see, for instance, \cite{BQ1_2000,Bag_2001,Ahm_2001,BQ2_2002,Wadati,Ortiz_2015}). The concept of \PT symmetry later spread to optics \cite{Christodoulides2008}, because the linear paraxial light propagation is also governed by the Schr\"odinger equation, where the real part of the potential is the refractive index, and the imaginary part of the potential describes the gain and loss in the medium. In the optical setting, \PT symmetry means that the refractive index is an even function in space, while the gain-loss profile is an odd function in space. In addition to optics, \PT symmetry has been introduced in other physical areas such as Bose-Einstein condensation \cite{BEC}. More importantly, many experimental observations and emerging applications of \PT symmetry have been reported \cite{Ruter_2010,Regensburger_2012,Feng2013,PTlaser_Zhang,PTlaser_CREOL,Peng}. For a review on this subject, see \cite{review_PT,Kivshar_review,PTbook_Yang,PTbook_Bender,review_Znojil}.

Generalization of \PT-symmetric potentials with real spectra is clearly an important issue. One would wonder, what non-\PT-symmetric complex potentials could still admit all-real spectra. Progress on this question has been made by three main methods. The first one is the supersymmetry (SUSY) method. SUSY was originally developed to find real partner potentials which share the same spectrum as the base real potential \cite{Infeld_1951,Wit_1981,Cooper_1995}. This technique starts with the factorization of a Schr\"odinger operator with a base potential into the product of two first-order operators. Switching the order of these two operators gives another Schr\"odinger operator with a partner potential, which shares the same spectrum as the base potential (except possibly a single discrete eigenvalue). Combining this idea of SUSY with the fact that the factorization of a Schr\"odinger operator is not unique \cite{Mlk_1984, Fer_1984, Mit_1989, Khare89}, parametric families of non-\PT-symmetric complex potentials with all-real spectra can be constructed \cite{Cannata98, Miri2013, Yang2014}. Note that in all such constructions in the past, an eigenmode of the base potential was always utilized. The second method is based on pseudo-Hermiticity \cite{Mostafazadeh1}. An operator is said to be pseudo-Hermitian if it is related to its Hermitian through a similarity transformation. As an example, all \PT-symmetric operators are pseudo-Hermitian. For a pseudo-Hermitian operator, its complex eigenvalues always appear as conjugate pairs. This conjugate-pair eigenvalue symmetry does not guarantee an all-real spectrum; but in many cases, it does force the spectrum to be all-real, as we have seen in many \PT-symmetric potentials. If certain conditions on the similarity transformation are further satisfied, then the spectrum will be all-real \cite{Mostafazadeh2}. Using ideas of pseudo-Hermiticity, wide classes of non-\PT-symmetric potentials with all-real spectra were identified \cite{BQ3_2002, NY, Yang2017}. The third method is based on the connection between the Schr\"odinger eigenvalue problem and the Zakharov-Shabat eigenvalue problem if the potential of the Schr\"odinger operator is of the form
$V(x)=u^2(x)+iu'(x)$, where $u(x)$ is a real function \cite{Wadati,Tsoy}, and the prime represents the derivative. This form of the potential is sometimes referred to as the Wadati potential. Then, utilizing available information on the Zakharov-Shabat eigenvalue problem and the soliton theory, all-real spectrum for the potential of the above form could be established if $u(x)$ is a $N$-soliton solution of the modified Korteweg-de Vries equation \cite{Wadati}, or any single-humped localized real function \cite{Tsoy,Shaw}. If such $u(x)$ is not even, then the resulting complex potential $V(x)$ would be non-\PT-symmetric. A review on these various methods can be found in \cite{YA}.

Despite this progress, construction of wider classes of non-\PT-symmetric complex potentials with all-real spectra is still an important endeavor, since such potentials are significant in diverse physical fields as already noted. In this article, we construct new families of non-\PT-symmetric complex potentials with all-real spectra by a combination of SUSY and the pseudo-Hermiticity method. Different from previous SUSY derivations of complex potentials, our treatment does not utilize eigenmodes of base potentials. As a result, our new potentials feature explicit analytical expressions which contain free functions. With the pseudo-Hermiticity method, we derive a new class of non-\PT-symmetric complex potentials with free functions and constants, whose eigenvalues appear as conjugate pairs. This eigenvalue symmetry constrains the spectrum to be all-real for a wide range of choices of these functions and constants in the potential. Tuning these free functions and constants, phase transition can also be induced, where conjugate pairs of complex eigenvalues emerge in the spectrum. Numerical examples of these new non-\PT-symmetric complex potentials and their spectra are also illustrated.

\section{Explicit non-\PT-symmetric potentials with real spectra by SUSY}
In this section, we construct non-\PT-symmetric potentials with real spectra by the SUSY method. Our main result is the following.

\textbf{Proposition 1.} The two potentials
\begin{equation} \label{eq:V0}
V_0=h'(x)-h^2(x)
\end{equation}
and
\begin{equation} \label{eq:V}
V=-h'(x)-h^2(x) +2\frac{d^2}{dx^2}\ln\left[c+\int_{0}^x e^{2\int_{0}^\xi h(\eta)d\eta}d\xi\right]
\end{equation}
are partner potentials which are related by
\begin{eqnarray}
&& -\partial_{xx}-V_0=(-\partial_x+W)(\partial_x+W), \label{fac1} \\
&& -\partial_{xx}-V=(\partial_x+W)(-\partial_x+W),   \label{fac2}
\end{eqnarray}
where $h(x)$ is an arbitrary complex function, $c$ is an arbitrary complex constant, and
\begin{equation} \label{e:W}
W(x)=h(x)-\frac{d}{dx}\ln\left[c+\int_{0}^x e^{2\int_{0}^\xi h(\eta)d\eta}d\xi\right].
\end{equation}

\textbf{Remark 1.} Since $V_0$ and $V$ above are partner potentials related through (\ref{fac1})-(\ref{fac2}), if $\psi(x)$ is a discrete eigenfunction of $V_0$ with eigenvalue $\lambda$, then $(\partial_x+W)\psi(x)$ would be a discrete eigenfunction of $V$ with the same eigenvalue $\lambda$ if $(\partial_x+W)\psi(x)$ is localized and not a zero function. Likewise, if $\phi(x)$ is a discrete eigenfunction of $V$ with eigenvalue $\omega$, then $(-\partial_x+W)\phi(x)$ would be a discrete eigenfunction of $V_0$ with the same eigenvalue $\omega$ if $(-\partial_x+W)\phi(x)$ is localized and not a zero function. In generic cases, this means that the two potentials $V_0$ and $V$ share exactly the same spectrum, as examples in later texts will illustrate. This also means that generically, if $V_0$ admits an exceptional point in the spectrum (by colliding two real eigenvalues for instance), then its partner potential $V$ would admit this same exceptional point as well.

\textbf{Remark 2.} Compared to previous constructions of complex potentials by the SUSY method, a distinctive feature of our result in Proposition 1 is that our potential $V(x)$ in (\ref{eq:V}) does not utilize discrete eigenmodes of the base potential $V_0(x)$. It only uses the function $h(x)$ which appears in the expression of the base potential. The advantage of this is that, for a given function $h(x)$, the $V(x)$ expression (\ref{eq:V}) is completely explicit. If the spectrum of $V_0(x)$ in (\ref{eq:V0}) is known, then the spectrum of this explicit potential $V(x)$ will be known as well for an arbitrary complex constant $c$. For instance, if $h(x)$ is chosen as an arbitrary real function, then the potential $V(x)$ from (\ref{eq:V}) with any complex constant $c$, which is non-\PT-symmetric in general, would have all-real spectrum since the spectrum of the corresponding real base potential $V_0(x)$ is all-real. A more important thing is, recent progress shows that, for wide choices of complex functions $h(x)$, the spectrum of the complex base potential (\ref{eq:V0}) is all-real as well. Examples include the case where $h(x)=ig(x)$, with $g(x)$ being an arbitrary real function (the so-called Wadati potential) \cite{BQ3_2002,Wadati,Tsoy,NY}, and the case where $h(x)$ is an arbitrary \PT-symmetric function \cite{Yang2017}. For these choices of the $h(x)$ function, we would be able to explicitly construct their partner potentials $V(x)$ from (\ref{eq:V}) which feature all-real spectra.

\textbf{Remark 3.} If we let $h(x)=ig(x)$, with $g(x)$ being a generally-complex function satisfying the asymptotics of $g(x) \to \mp k_0$ as $x\to \pm \infty$ and $k_0$ being a non-zero real constant, then the potential $V_0(x)$ in Eq. (\ref{eq:V0}), i.e., $V_0=g^2(x)+ig'(x)$, would
admit a spectral singularity \cite{KonotopSS}, and the associated wavefunction at this spectral singularity is $\psi(x)=\exp\left[-i\int_0^x g(\xi)d\xi\right]$, which features outgoing plane wave asymptotics as $x\to \pm \infty$. However, our new potential $V(x)$ in Eq. (\ref{eq:V}) would not inherit this spectral singularity from its partner potential $V_0$ in general, because the transformed wavefunction $(\partial_x+W)\psi(x)$ for the new potential $V(x)$ generically would not exhibit outgoing plane wave asymptotics as $x\to \pm \infty$.

\textbf{Proof of Proposition 1.} For the base potential $V_0$ in (\ref{eq:V0}), the Schr\"odinger operator can be factorized as
\[
-\partial_{xx}-V_0=(-\partial_x+h)(\partial_x+h).
\]
It is important to realize that this factorization is not unique \cite{Mlk_1984, Fer_1984, Mit_1989, Khare89}. Indeed, it is easy to check that this same Schr\"odinger operator can also be factorized as
\[ \label{fac4}
-\partial_{xx}-V_0=(-\partial_x+W)(\partial_x+W),
\]
where $W(x)$ is related to $h(x)$ by Eq. (\ref{e:W}). This $W(x)$ can also be derived judiciously by equating the right sides of the above two factorizations, which results in
\[
W'-W^2=h'-h^2.
\]
This is a Riccati equation for $W(x)$. Employing the variable transformation $W=h+w^{-1}$, the equation for $w$ reduces to a linear equation $w'+2hw =-1$. Solving this $w(x)$ equation and putting results together, the $W(x)$ formula (\ref{e:W}) would be derived.

Based on the second factorization (\ref{fac4}), we construct the partner potential $V(x)$ through
\[
-\partial_{xx}-V=(\partial_x+W)(-\partial_x+W),
\]
i.e., $V=-W'-W^2$. Inserting the $W(x)$ formula (\ref{e:W}) and simplifying, the expression (\ref{eq:V}) for $V(x)$ would be obtained. This finishes the proof of Proposition 1.

Next, we use three examples to illustrate the applications of Proposition 1. Notation-wise, an eigenvalue $\lambda$ of a potential $V(x)$ is defined by
\[
\left[\partial_{xx}+V(x)\right]\psi(x)=\lambda\psi(x).
\]

\textbf{Example 1.} In our first example, we choose $h(x)$ to be a \PT-symmetric function. The reason for it is that when $h(x)$ is \PT-symmetric, the spectrum of the base potential (\ref{eq:V0}) features conjugate-pair eigenvalue symmetry; hence its spectrum is often all-real \cite{Yang2017}. As a result, the spectrum of the non-\PT-symmetric partner potential (\ref{eq:V}) would also be all-real for an arbitrary complex constant $c$. To be concrete, we choose $h(x)$ and $c$ to be the following \PT-symmetric function and constant
\[ \label{h1}
h(x)=\mbox{sech}^2 x+i \hspace{0.05cm} \mbox{sech} x \hspace{0.05cm} \tanh x, \quad c=3e^{2i}.
\]
The resulting base and partner potentials (\ref{eq:V0}) and (\ref{eq:V}) are displayed in Fig. 1(a,b) respectively. Notice that both potentials are non-\PT-symmetric. The spectra of these two potentials are displayed in Fig. 1(c,d). Not surprisingly, the spectrum of $V_0$ is all-real. Then, based on Proposition 1, $V$ and $V_0$ share the same real spectrum, as Fig. 1(d) confirms. Notice that these spectra do not contain any discrete real eigenvalues. This is a general property of the base potential (\ref{eq:V0}) when $h(x)$ is a localized \PT-symmetric function
\cite{Yang2017}.

\begin{figure}[htbp]
\begin{center}
\includegraphics[width=0.5\textwidth]{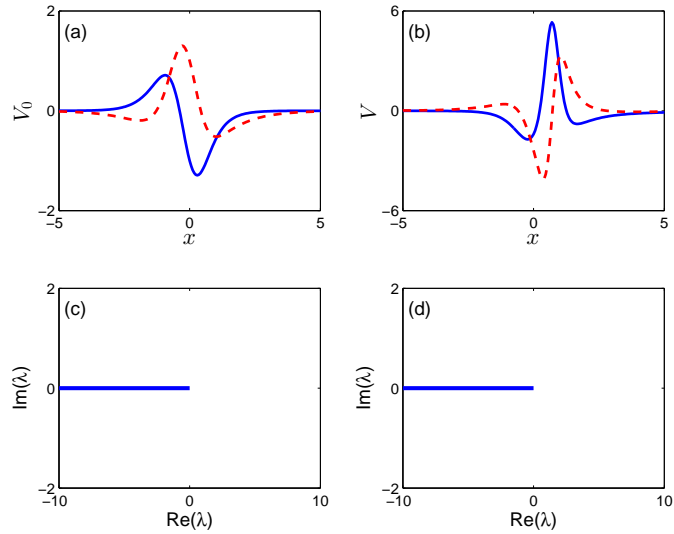}
\caption{Spectra of the base and partner potentials $V_0(x)$ and $V(x)$ with choices of $h(x)$ and $c$ given in Eq. (\ref{h1}).
(a) Profile of $V_0$. (b) Profile of $V$. (c) Spectrum of $V_0$. (d) Spectrum of $V$. The solid blue and dashed red lines in (a, b) represent the real and imaginary parts of the potential respectively. } \label{f:fig1}
\end{center}
\end{figure}

\textbf{Example 2.} In our second example, we choose $h(x)=ig(x)$, where $g(x)$ is an arbitrary real function. This choice leads to $V_0=g^2(x)+ig'(x)$, which is the so-called Wadati potential \cite{Wadati}. This potential is generally non-\PT-symmetric, but its spectrum is often all-real \cite{BQ3_2002,Wadati,Tsoy,NY,YA}. As a consequence, the spectrum of its partner potential (\ref{eq:V}) is also all-real. To be concrete, we choose $h(x)$ and $c$ as
\[ \label{h2}
h(x)=i\left[\mbox{sech}^2(x+1)+1.5\mbox{sech}^2(x-1)\right], \quad c=2e^{i\hspace{0.03cm}\tanh(1)}.
\]
The resulting base and partner potentials are displayed in Fig. 2(a,b), followed by their spectra in Fig. 2(c,d), respectively. In this case, the spectrum of $V_0$ is all-real with one discrete eigenvalue. The spectrum of the non-\PT-symmetric partner potential $V$ is the same as that for $V_0$, as Fig. 2(d) confirms.

\begin{figure}[htbp]
\begin{center}
\includegraphics[width=0.5\textwidth]{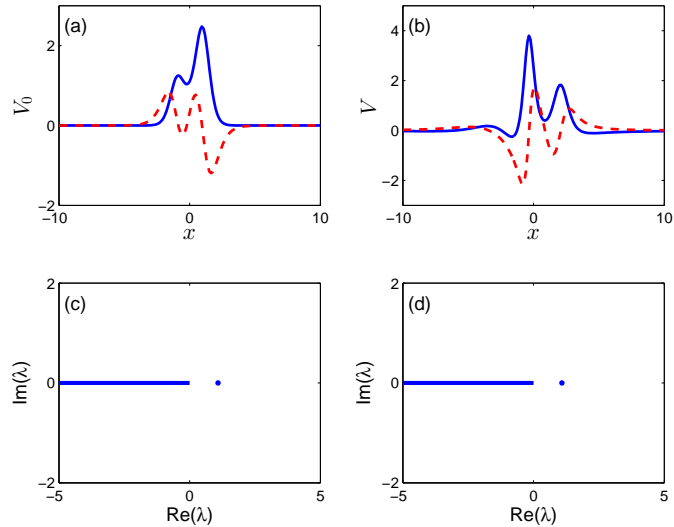}
\caption{Spectra of the base and partner potentials $V_0(x)$ and $V(x)$ with choices of $h(x)$ and $c$ given in Eq. (\ref{h2}).
(a) Profile of $V_0$. (b) Profile of $V$. (c) Spectrum of $V_0$. (d) Spectrum of $V$.} \label{f:fig2}
\end{center}
\end{figure}

\textbf{Example 3.} In our third example, we choose $h(x)$ to be an arbitrary real function. In this case, the base potential $V_0$ is real; thus its spectrum is all-real. Proposition 1 then indicates that the spectrum of the partner potential $V(x)$ would be all-real as well. To illustrate, we choose $h(x)$ and $c$ as
\[ \label{h3}
h(x)=x \hspace{0.06cm} e^{-x^2/2}, \quad c=(5+i)e^{2}.
\]
The resulting potentials $V_0$, $V$ and their spectra are displayed in Fig. 3. In this case, the spectrum of $V_0$ is all-real with no discrete eigenvalue. The spectrum of the non-\PT-symmetric partner potential $V$ is the same as that for $V_0$, as Fig. 3(d) confirms.

\begin{figure}[htbp]
\begin{center}
\includegraphics[width=0.5\textwidth]{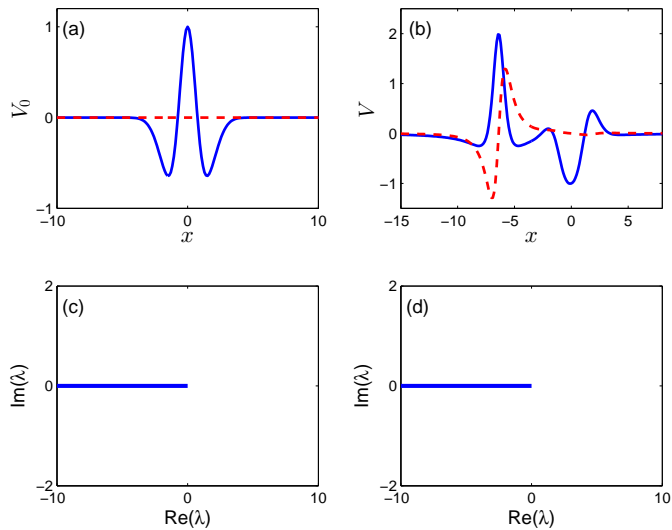}
\caption{Spectra of the base and partner potentials $V_0(x)$ and $V(x)$ with choices of $h(x)$ and $c$ given in Eq. (\ref{h3}).
(a) Profile of $V_0$. (b) Profile of $V$. (c) Spectrum of $V_0$. (d) Spectrum of $V$.} \label{f:fig3}
\end{center}
\end{figure}

\section{A family of parity-induced non-\PT-symmetric potentials by the pseudo-Hermiticity method}
The pseudo-Hermiticity method is based on a pseudo-Hermitian-like relation \cite{Mostafazadeh1}
\[ \label{etaL}
\eta L=L^\dagger \eta,
\]
where
\[
L=\partial_{xx}+V(x)
\]
is a Schr\"odinger operator, $V(x)$ is a complex potential, $L^\dagger=\partial_{xx}+V^*(x)$ is the Hermitian of $L$, the superscript `*' represents complex conjugation, and $\eta$ is another operator. But unlike pseudo-Hermiticity, $\eta$ is not required to be invertible here. If a complex potential satisfies the above relation, then when the kernel of $\eta$ is empty, complex eigenvalues of $L$ would come as conjugate pairs \cite{NY}. This conjugate-pair eigenvalue symmetry often leads to all-real spectra, but phase transition can also occur \cite{NY,Yang2017}, where conjugate pairs of complex eigenvalues appear in the spectrum, just like in many \PT-symmetric potentials \cite{Bender1998,Ahm_2001,Christodoulides2008}. When $\eta$ is chosen as a differential operator, then several families of non-\PT-symmetric potentials satisfying the above symmetry relation have been derived, with the Wadati potential as a particular example \cite{NY}. If $\eta$ is chosen as a combination of the parity operator and a first-order differential operator, then the resulting family of potentials is $V=h'(x)-h^2(x)$, where $h(x)$ is a \PT-symmetric function \cite{YA}.

In this section, we generalize the derivation in \cite{YA} by choosing $\eta$ to be a combination of the parity operator and a second-order differential operator. As we will see, this will produce a new family of non-\PT-symmetric potentials admitting the pseudo-Hermitian relation
(\ref{etaL}), hence opening the door for all-real spectra for a wide range of potentials in this family.

We start by postulating $\eta$ as
\begin{equation}
\eta = {\cal{P}} [\partial_{xx} + a(x)\partial_x + b(x)],
\end{equation}
where $\cal{P}$ is the parity operator, i.e., ${\cal{P}}f(x)\equiv f(-x)$, and the complex functions $a(x)$, $b(x)$ are to be determined. Substituting this $\eta$ and $L$, $L^\dagger$ into Eq. (\ref{etaL}) and collecting terms of the same order of derivatives on the two sides, we get the following series of equations,
\begin{equation}  \label{e:eq1}
V(x) - V^*(-x) =2a',
\end{equation}
\begin{equation}  \label{e:eq2}
a[V(x) - V^*(-x)] = a'' +2 b' -2V'(x),
\end{equation}
\begin{equation} \label{e:eq3}
b[V(x) - V^*(-x)]= b''-V''(x) -aV'(x).
\end{equation}
From the first equation, we see that $a'(x) = [a^*(-x)]_x$. Hence we can choose $a(x)$ such that
\[
a^*(-x)=a(x),
\end{equation}
i.e., $a(x)$ is \PT-symmetric. When the first equation (\ref{e:eq1}) is inserted into the second equation (\ref{e:eq2}) and integrating once, we get
\begin{equation}
b = \frac{1}{2} (a^2 -a') +V-c_1,
\end{equation}
where $c_1$ is an integrating constant. Utilizing this $b$ expression as well as Eq. (\ref{e:eq1}), we find that Eq. (\ref{e:eq3}) becomes
\begin{equation}
\left(a^2V\right)' = \left(c_1a^2+\frac{1}{4} (a'^2 -2a'' a)+a^2a'-\frac{1}{4}a^4\right)'.
\end{equation}
By integrating this equation, we obtain an explicit expression for the potential $V(x)$. Neglecting an overall constant $c_1$, this potential reads
\begin{equation} \label{e:VII}
V(x) = a' -\frac{1}{4} a^2 + \frac{a'^2 -2a'' a +c_2}{4a^2},
\end{equation}
where $c_2$ is a constant of integration. In view that $a(x)$ is \PT-symmetric, in order for the above $V(x)$ function to satisfy Eq. (\ref{e:eq1}), $c_2$ must be real.

Thus, we have derived a new family of generally non-\PT-symmetric potentials (\ref{e:VII}), where $a(x)$ is an arbitrary \PT-symmetric function, and $c_2$ is an arbitrary real constant. This family of potentials satisfy the pseudo-Hermiticity relation (\ref{etaL}), and hence its eigenvalues appear as complex-conjugate pairs. This conjugate-pair eigenvalue symmetry does not guarantee a real spectrum, but it does often force the spectrum to be all-real, similar to \PT-symmetric potentials. Notice that the analytical form (\ref{e:VII}) for this family of potentials bears similarity to type-II potentials derived in \cite{NY} when $\eta$ was taken as a second-order differential operator without the inclusion of the parity operator. Of course, important differences exist between these two families of potentials as well.

Next, we use an example to illustrate the spectral properties of this family of potentials. For this purpose, we choose
\[ \label{ab}
a(x)=\mbox{sech} x+i\hspace{0.03cm} \beta \hspace{0.04cm} \mbox{sech}x\tanh \hspace{-0.03cm}x-2, \quad c_2=3,
\]
where $\beta$ is a real parameter. Notice that $a(x)$ is \PT-symmetric and $c_2$ real as required. When $\beta=1$, the resulting potential (\ref{e:VII}) and its spectrum are displayed in Fig. 4(a, c) respectively. It is seen that for this non-\PT-symmetric potential, its spectrum is all-real with two discrete eigenvalues. But when $\beta$ increases to 2, whose potential is plotted in Fig. 4(b), we see from Fig. 4(d) that a pair of complex eigenvalues appear in the spectrum. In other words, phase transition has occurred.

\begin{figure}[htbp]
\begin{center}
\includegraphics[width=0.5\textwidth]{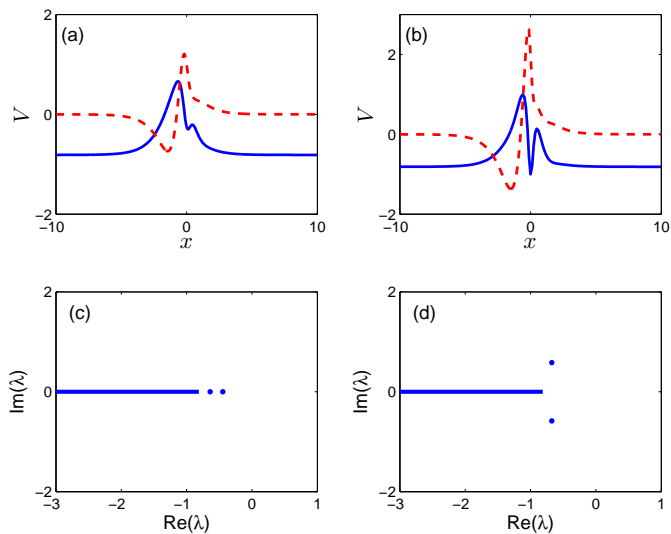}
\caption{Spectra of potentials (\ref{e:VII}) where $a(x)$ and $c_2$ are chosen as in Eq. (\ref{ab}).
(a) Profile of $V$ when $\beta=1$. (b) Profile of $V$ when $\beta=2$. (c) Spectrum of the potential in (a). (d) Spectrum of the potential in (b).} \label{f:fig4}
\end{center}
\end{figure}

\section{Summary}

Two methods, one based on the idea of SUSY and the other exploiting the pseudo-Hermiticity principle, were examined to uncover new families of non-\PT-symmetric complex potentials that support all-real spectra. The SUSY method that we explored differs from the previous attempts in the sense that it avoids using discrete eigenmodes of base potentials. As a result, the analytic expressions of our non-\PT-symmetric partner potentials are explicit with free functions, which is an advantage when constructing such potentials for all-real spectra.
With the pseudo-Hermiticity method and a suitable choice of the connecting operator as a combination of parity and a second-order differential operator, a class of non-\PT-symmetric potentials with conjugate-pair eigenvalue symmetry were derived. Numerical examples have also been given to illustrate the spectral properties of the potentials that we produced.

\section*{Acknowledgment}
One of us (BB) gratefully acknowledges the warm hospitality of Prof. J. Yang at the University of Vermont, where this work was initiated. This material is based upon work supported by the Air Force Office of Scientific Research under award number FA9550-18-1-0098, and the National Science Foundation under award number DMS-1616122.

\section*{Data Availability Statement}
The data that supports the findings of this study are available within the article.


\vspace{0.3cm}
\begin{thebibliography}{10}

\bibitem{Infeld_1951}
L. Infeld and T. E. Hull,
``The Factorization Method",
Rev. Mod. Phys. 23, 21 (1951).

\bibitem{Bhatt_1962}
A. Bhattacharjie and E. C. G. Sudarshan,
``A class of solvable potentials",
Nuovo Cimento 25, 864 (1962).

\bibitem{Levai_1989}
G. Levai,
``On some exactly solvable potentials derived from supersymmetric quantum mechanics",
J. Phys. A 22, 689 (1989).

\bibitem{Cooper_1995}
F. Cooper, A. Khare and U. Sukhatme,
``Supersymmetry and quantum mechanics",
Phys. Rep. 251, 267 (1995).

\bibitem{Bender1998}
C. M. Bender and S. Boettcher,
``Real spectra in non-Hermitian Hamiltonians having \PT symmetry",
Phys. Rev. Lett. 80, 5243 (1998).

\bibitem{Mostafazadeh_quantum}
A. Mostafazadeh, ``Pseudo-Hermitian representation of quantum mechanics", Int. J. Geom.
Methods Mod. Phys. 7, 1191–1306 (2010).

\bibitem{Moise_2011}
N. Moiseyev, \emph{Non-Hermitian Quantum Mechanics} (Cambridge University Press, Cambridge, 2011).

\bibitem{Brody_2016}
D.C. Brody,
``Consistency of PT-symmetric quantum mechanics",
J. Phys. A 49, 1751 (2016).

\bibitem{BQ1_2000}
B. Bagchi and C. Quesne,
``sl(2, C) as a complex Lie algebra and the associated non-Hermitian Hamiltonians with real eigenvalues",
Phys. Lett. A 273, 285 (2000).

\bibitem{Bag_2001}
B. Bagchi, S. Mallik and C. Quesne,
``Generating Complex Potentials with Real Eigenvalues in Supersymmetric Quantum Mechanics",
Int. J. Mod. Phys. A 16, 2859 (2001).

\bibitem{Ahm_2001}
Z. Ahmed,
``Real and Complex discrete eigenvalues in an exactly solvable one-dimensional complex PT-invariant potential",
Phys. Lett. A 282, 343 (2001).

\bibitem{BQ2_2002}
B. Bagchi and C. Quesne,
``Non-Hermitian Hamiltonians with real and complex eigenvalues in a Lie-algebraic framework",
Phys. Lett. A 300, 18 (2002).

\bibitem{Wadati} M. Wadati,
``Construction of parity-time symmetric potential through the soliton theory",
J. Phys. Soc. Jpn. 77, 074005 (2008).

\bibitem{Ortiz_2015}
O. Rosas-Ortiz, O. Castanos and D. Schuch,
``New supersymmetry-generated complex potentials with real spectra",
J. Phys. A 48, 445302 (2015).

\bibitem{Christodoulides2008}
Z.H. Musslimani, K.G. Makris, R. El-Ganainy and D.N. Christodoulides, ``Optical solitons in
\PT periodic potentials", Phys. Rev. Lett. 100, 030402 (2008).

\bibitem{BEC}
S. Klaiman, U. G\"unther and N. Moiseyev, ``Visualization of branch points in \PT-symmetric waveguides", Phys. Rev. Lett.
101, 080402 (2008).

\bibitem{Ruter_2010}  C.E. R\"uter, K.G. Makris, R. El-Ganainy, D.N. Christodoulides, M. Segev and D. Kip,
``Observation of parity–-time symmetry in optics", Nature Physics 6, 192--195 (2010).

\bibitem{Regensburger_2012}
A. Regensburger, C. Bersch, M.A. Miri, G. Onishchukov, D.N. Christodoulides and U. Peschel,
``Parity–time synthetic photonic lattices", Nature 488, 167--171 (2012).

\bibitem{Feng2013}
L. Feng, Y.L. Xu,  W.S. Fegadolli, M.H. Lu, J.E.B. Oliveira, V.R. Almeida, Y.F. Chen, and A. Scherer,
``Experimental demonstration of a unidirectional reflectionless parity-time metamaterial at optical frequencies",
Nature Materials, 12, 108--113 (2013).

\bibitem{PTlaser_Zhang} L. Feng, Z.J. Wong, R. Ma, Y. Wang, and X. Zhang,
``Single-mode laser by parity-time symmetry breaking",
Science 346, 972--975 (2014).

\bibitem{PTlaser_CREOL} H. Hodaei, M.-A. Miri, M. Heinrich, D. N. Christodoulides and M. Khajavikhan,
``\PT-symmetric micro-ring laser",
Science 346, 975--978 (2014).

\bibitem{Peng}
B. Peng, S.K. \"Ozdemir, F. Lei, F. Monifi, M. Gianfreda, G.L. Long,
S. Fan, F. Nori, C.M. Bender and L. Yang, ``Parity-time-symmetric whispering-gallery microcavities", Nat. Phys. 10,
394--398 (2014).

\bibitem{review_PT}
V.V. Konotop, J. Yang and D.A. Zezyulin, ``Nonlinear waves in \PT-symmetric systems", Rev. Mod.
Phys. 88, 035002 (2016).

\bibitem{Kivshar_review}
S.V. Suchkov, A.A. Sukhorukov, J. Huang, S.V. Dmitriev, C. Lee and Y.S. Kivshar,
``Nonlinear switching and solitons in \PT-symmetric photonic systems",
Laser Photon. Rev. 10, 177 (2016).


\bibitem{PTbook_Yang}
D.N. Christodoulides and J. Yang (Eds), \emph{Parity-time Symmetry and Its Applications} (Springer, Singapore, 2018).

\bibitem{PTbook_Bender}
C.M. Bender, \emph{\PT Symmetry in Quantum and Classical Physics} (World Scientific, Singapore, 2019).

\bibitem{review_Znojil}
F. Bagarello, J.P. Gazeau, F.H. Szafraniec and M. Znojil (Eds),
\emph{Non-Self-Adjoint Operators in Quantum Physics: Mathematical Aspects} (Wiley, 2015).

\bibitem{Wit_1981}
E. Witten,
``Dynamical breaking of supersymmetry",
Nucl. Phys. B 188, 513 (1981).

\bibitem{Mlk_1984}
B. Mielnik,
``Factorization method and new potentials with the oscillator spectrum",
J. Math. Phys. 25, 3387 (1984).

\bibitem{Fer_1984}
D. J. Fernandez,
``New hydrogen-like potentials",
Lett. Math. Phys. 8, 337 (1984).

\bibitem{Mit_1989}
A. Mitra, A. Lahiri, P. K. Roy and B. Bagchi,
``Nonuniqueness of the factorization scheme in quantum mechanics",
Int. J. Theor. Phys. 28, 911 (1989).

\bibitem{Khare89}
A. Khare and U. Sukhatme, ``Phase-equivalent potentials obtained from
supersymmetry", J. Phys. A 22, 2847--2860 (1989).

\bibitem{Cannata98}
F. Cannata, G. Junker and J. Trost,
``Schr\"odinger operators with complex potential but real spectrum",
Phys. Lett. A 246, 219--226 (1998).

\bibitem{Miri2013}
M. Miri, M. Heinrich and D.N. Christodoulides,
``Supersymmetry-generated complex optical potentials with real spectra",
Phys. Rev. A 87, 043819 (2013).

\bibitem{Yang2014}
J. Yang, ``Necessity of \PT symmetry for soliton families in one-dimensional complex potentials", Phys. Lett. A 378, 367--373 (2014).

\bibitem{Mostafazadeh1}
A. Mostafazadeh, ``Pseudo-Hermiticity versus \PT symmetry: The necessary condition for the reality of the
spectrum of a non-Hermitian Hamiltonian", J. Math. Phys. 43, 205 (2002).

\bibitem{Mostafazadeh2}
A. Mostafazadeh,
``Pseudo-Hermiticity versus \PT-symmetry. II. A complete characterization of non-Hermitian Hamiltonians
with a real spectrum", J. Math. Phys. 43, 2814 (2002).

\bibitem{BQ3_2002}
B. Bagchi and C. Quesne,
``Pseudo-Hermiticity, weak pseudo-Hermiticity and eta-orthogonailty condition",
Phys. Lett. A 301, 173 (2002).

\bibitem{NY} S. Nixon and J. Yang, ``All-real spectra in optical systems with arbitrary gain-and-loss distributions",
Phys. Rev. A 93, 031802(R) (2016).

\bibitem{Yang2017}
J. Yang, ``Classes of non-parity-time-symmetric optical potentials with exceptional-point-free phase transitions", Opt. Lett. 42, 4067 (2017).

\bibitem{Tsoy} E.N. Tsoy, I.M. Allayarov and F. Kh. Abdullaev,
``Stable localized modes in asymmetric waveguides with gain and loss",
Opt. Lett. 39, 4215 (2014).

\bibitem{Shaw} M. Klaus and J. K. Shaw, ``Purely imaginary eigenvalues of Zakharov-Shabat systems", Phys. Rev. E 65, 036607 (2002).

\bibitem{YA} J. Yang, ``Construction of non-\PT-symmetric complex potentials with all-real spectra", book chapter in
\emph{Parity-Time Symmetry and Its Applications} (D. Christodoulides and J. Yang (Eds), Springer, Singapore), pp. 513--534 (2018).

\bibitem{KonotopSS} D.A. Zezyulin and V.V. Konotop, ``A universal form of localized complex potentials with spectral singularities",  New J. Phys. 22 013057 (2020).

\end{thebibliography}
\end{document}